\documentclass{article}

\usepackage{arxiv}
\usepackage{natbib}
\usepackage{subcaption}
\usepackage{float}
\usepackage[utf8]{inputenc} 
\usepackage[T1]{fontenc}    
\usepackage{hyperref}       
\usepackage{url}            
\usepackage{booktabs}       
\usepackage{amsfonts}       
\usepackage{nicefrac}       
\usepackage{microtype}      
\usepackage{lipsum}
\usepackage{graphicx}
\usepackage{comment}
\graphicspath{ {./images/} }

\title{An Empirical Analysis of Zero-Day Vulnerabilities Disclosed by the Zero Day Initiative (January–April 2024)}

\author{
 Apurva M Shet \\
  Department of Computing, Engineering and Mathematical Sciences\\
  Texas A\&M University - San Antonio\\
  San Antonio, TX 78224 \\
  \texttt{ashet02@jaguar.tamu.edu} \\
   \And
 Izzat Alsmadi \\
  Department of Computing, Engineering and Mathematical Sciences\\
  Texas A\&M University - San Antonio\\
  San Antonio, TX 78224\\
  \texttt{ialsmadi@tamusa.edu} \\
}

\begin{document}
\maketitle
\begin{abstract}
Zero-day vulnerabilities represent some of the most critical threats in cybersecurity, as they correspond to previously unknown flaws in software or hardware that are actively exploited before vendors can develop and deploy patches. During this exposure window, affected systems remain defenseless, making zero-day attacks particularly damaging and difficult to mitigate. This study analyzes the Zero Day Initiative (ZDI) vulnerability disclosures reported between January and April 2024, \cite{zdikaggle2025} comprising a total of 415 vulnerabilities. The dataset includes vulnerability identifiers, Common Vulnerability Scoring System (CVSS) v3.0 scores, publication dates, and short textual descriptions. The primary objectives of this work are to identify trends in zero-day vulnerability disclosures, examine severity distributions across vendors, and investigate which vulnerability characteristics are most indicative of high severity. In addition, this study explores predictive modeling approaches for severity classification, comparing classical machine learning techniques with deep learning models using both structured metadata and unstructured textual descriptions. The findings aim to support improved patch prioritization strategies, more effective vulnerability management, and enhanced organizational preparedness against emerging zero-day threats.
\end{abstract}


\section{Introduction} Zero-day vulnerabilities are software or hardware flaws that are exploited by adversaries before the responsible vendor becomes aware of their existence or releases a corresponding patch. Because no official mitigation is available at the time of exploitation, zero-day attacks are particularly dangerous and often result in successful compromises of high-value targets \cite{roumani2021patching,guo2023review}. A zero-day exploit refers to the active use of such a vulnerability, typically occurring immediately upon or even before public disclosure, leaving defenders with minimal response time.

To standardize the assessment of vulnerability severity, the cybersecurity community relies on the Common Vulnerability Scoring System (CVSS). CVSS provides a numerical score ranging from 0 to 10 that reflects the potential impact and exploitability of a vulnerability. While the CVSS Base score is most commonly used for prioritization, Temporal and Environmental metrics further account for exploit maturity, remediation availability, and organizational context \cite{hoque2021improved,iannone2024early}. Despite its widespread adoption, CVSS-based prioritization remains challenging due to the sheer volume of disclosures and the limited resources available for remediation.

Effective vulnerability management also depends on consistent identification mechanisms. The Common Vulnerabilities and Exposures (CVE) system assigns unique identifiers to publicly disclosed vulnerabilities, enabling standardized communication across vendors, researchers, and security practitioners. However, many zero-day vulnerabilities are initially disclosed without CVE identifiers, complicating early-stage risk assessment.

Private vulnerability disclosure programs play a critical role in bridging this gap. The Zero Day Initiative (ZDI) incentivizes independent researchers to responsibly disclose vulnerabilities to vendors, facilitating coordinated remediation efforts prior to public release \cite{roumani2021patching}. The dataset analyzed in this study consists of 415 ZDI-reported vulnerabilities disclosed between January and April 2024 \cite{zdikaggle2025}. Beyond descriptive analysis, this work explores features predictive of vulnerability severity and evaluates both classical and deep learning models for severity classification, with particular attention to evaluation metrics suited for imbalanced datasets.
\section{Research Questions} This study is guided by the following research questions:

\subsection{RQ1: Feature Predictiveness and Dimensionality Reduction} Which features in the ZDI dataset---including affected vendor, CVE assignment status, and keywords within vulnerability descriptions (e.g., \emph{buffer overflow}, \emph{remote code execution})---are most predictive of high-severity vulnerabilities ($CVSS v3.0 \geq 7.0 \geq7.0$)? Furthermore, how do dimensionality reduction techniques such as TF--IDF for textual features and Principal Component Analysis (PCA) or Truncated Singular Value Decomposition (SVD) for high-dimensional feature spaces affect classification performance and interpretability?

We hypothesize that vulnerabilities affecting widely deployed vendors and those containing high-risk exploit-related terminology are more strongly associated with elevated CVSS scores. Additionally, dimensionality reduction is expected to reduce noise and improve the effectiveness of classical machine learning models.

\subsection{RQ2: Classical Versus Deep Learning Models} How do classical machine learning models---such as Logistic Regression, Decision Trees, and Random Forests---compare with deep learning approaches, including recurrent neural networks and Transformer-based models (e.g., BERT), in predicting vulnerability severity categories (Low, Medium, High) from ZDI descriptions?

Prior work suggests that while classical models perform well on structured features, they struggle with unstructured textual data \cite{sara2023static,abri2019performance}. In contrast, deep learning models are better suited to capture semantic context in vulnerability descriptions, potentially yielding superior performance \cite{sarhan2023zeroshot,hindy2020utilising}.

\subsection{RQ3: Evaluation Metrics Under Class Imbalance} Given the imbalance inherent in vulnerability datasets, how do evaluation metrics such as precision, recall, F1-score, and ROC--AUC provide deeper insights into model performance compared to accuracy alone? We hypothesize that accuracy may obscure poor detection of critical vulnerabilities, whereas recall and ROC--AUC better reflect a model's ability to identify high-severity cases \cite{guo2023review,iannone2024early}.

\section{RELATED CODES}

1.	CVE Severity Classification
A dataset of over 169,000 CVE–CWE entries that match vulnerability descriptions with their respective severity classes from low to critical is used in this study. For natural language-based classification, the model sequence incorporates word-frequency analysis, textual preprocessing, and deep learning techniques like LSTM networks and Naïve Bayes. The results show that specific keywords such as "buffer overflow", "remote code execution", and "denial of service" are highly predictive of high or critical severity, with the LSTM achieving an F1-score near 0.84. On the other hand, ZDI expands on this concept of textual modeling. It incorporates structured variables such as vendors, dates, and identities in addition to TF-IDF and PCA for dimensionality reduction to provide a better balanced trade-off between interpretability and predictive performance. 

2.	CVE Severity Classification: a comparative study
The dataset used in this comparative study originates from the complete NIST CVE database, combining textual descriptions with structured attributes such as vendor counts, product information, logical operators in configurations, and reference metadata. The model ranges from traditional machine learning algorithms like Random Forest and XGBoost to deep architectures, including transformer based models for text and multilayer perceptrons for structured data. The findings validate the value of multimodal learning by showing that Transformer-based models achieved a macro F1-score of about 92 percent, while a hybrid MLP Transformer method enhanced performance to about 94 percent. In comparison, the ZDI dataset adopts a similar hybrid methodology that fuses textual and structured features but emphasizes dimensionality reduction and interpretability, reflecting a modern and explainable adaptation of this research design.

3.	CVSS\_Prediction
The dataset contains nearly 90,000 vulnerability records, each having CVSS scores, CWE codes, and categorical metrics such as access vector, authentication, and impact on confidentiality, integrity, and availability. In order to forecast CVSS scores as continuous values, the model applies Random Forest, XGBoost, and neural networks to TF-IDF representations of textual summaries with one-hot-encoded categorical variables. The results are highly accurate, achieving R² scores close to 0.9 and identifying impact-related metrics and specific keywords as the strongest predictors. The model In comparison, the ZDI dataset focuses on more recent 2024 vulnerabilities and applies similar prediction goals but introduces PCA-based dimensionality reduction to enhance interpretability while maintaining predictive strength.

4.	Cybersecurity Threat Analysis
The dataset used in this merges several CISA vulnerability bulletins from 2022, resulting in a unified collection of about 4,000 records containing CVE IDs, vendors, products, CWEs, CVSS scores, and severity ratings. The model is primarily statistical and exploratory, employing clustering, summarization, and visualization instead of predictive machine learning. The results indicate that most vulnerabilities were network-based, had low exploitation complexity, and displayed a median CVSS score of 8.8, underscoring the prevalence of critical issues. The ZDI dataset, in contrast, converts this descriptive research method into predictive modeling by identifying the characteristics most closely linked to high-risk vulnerabilities and forecasting severity classes using comparable qualities.

5.	Celosia - Zero-day attack detection demo
The dataset used in the Celosia demo is the N-Baiot IoT collection, which contains network traffic data from smart devices such as doorbells, webcams, and baby monitors. This data includes both benign and malicious samples generated by Gafgyt malware. The model consists of a deep autoencoder built with Keras, trained on regular traffic to recognize device behavior, and is then paired with logistic regression to classify anomalies based on hidden layer representations. In comparison, the ZDI dataset applies the same zero-day detection principle. However, it operates on textual and structured vulnerability data rather than raw network traffic, using predictive modeling to classify vulnerability severity instead of anomaly detection.

\begin{table*}[t]
\centering
\caption{Comparison of ZDI 2024 Analysis with Existing Kaggle Notebooks}
\label{tab:kaggle_comparison}
\resizebox{\textwidth}{!}{%
\begin{tabular}{p{2.5cm} p{3cm} p{3.5cm} p{3cm} p{3.5cm} p{4.5cm}}
\toprule
\textbf{Author / Notebook} & \textbf{Focus Area} & \textbf{Techniques \& Models} & \textbf{Dataset Used} & \textbf{Key Insights} & \textbf{Comparison with ZDI 2024 Work} \\
\midrule
\textbf{CVSS Prediction} & Focused on predicting CVSS scores using both text and numerical features. & Used TF-IDF for text processing and models like Random Forest, Logistic Regression, and Gradient Boosting. & CVE (Common Vulnerabilities and Exposures) (NIST NVD, 1999–2020) & Random Forest performed best with strong and balanced predictions. & our ZDI 2024 model follows a similar text-based setup but compares different feature selection methods (SVD, Chi$^2$, MI) using Logistic Regression for consistency and interpretability. \\
\midrule
\textbf{CVE Severity Classification (Comparative)} & Compared multiple algorithms for CVE severity classification. & Logistic Regression, SVM, Naïve Bayes, and Random Forest. & Common Vulnerabilities and Exposures (CVE) sourced from NVD (1999-2024) & Found that Logistic Regression and SVM offered high precision and stable performance. & our model uses Logistic Regression as well but includes dimensionality reduction and feature engineering, making it more explainable and data-driven. \\
\midrule
\textbf{CVE Severity Classification} & Classified vulnerabilities based purely on textual data. & TF-IDF + Random Forest Classifier. & CVE and CWE mapping Dataset (2021) (NVD 2002-2021) & Demonstrated that text-based features alone can predict severity effectively. & our analysis expands this by including structured features (like vendor and CVE flags), improving accuracy and interpretability. \\
\midrule
\textbf{Cybersecurity Threat Analysis} & Conducted exploratory data analysis (EDA) on cyber threat data. & Visualization, correlation analysis, and statistical summaries. & Cybersecurity Risk (2022 CISA Vulnerability) & Found that network-based vulnerabilities dominate, with an average CVSS score of $\sim$6.8. & our work builds on this by going beyond EDA to predict vulnerability severity, combining both descriptive and predictive insights. \\
\midrule
\textbf{Celosia: Zero-Day Attack Detection} & Developed a detection pipeline for zero-day exploits. & Feature engineering with Random Forest and XGBoost. & Benign dataset from N-BaIoT Dataset to Detect IoT Botnet Attacks & Hybrid detection improved precision and recall. & our work applies similar ideas but uses real ZDI 2024 cases, linking textual and structured data for a realistic and interpretable severity model. \\
\bottomrule
\end{tabular}%
}
\end{table*}

\section{RELATED WORKS} 
\begin{itemize}

\item Static Analysis based Malware Detection for Zeroday Attacks in Android Applications: This study looked at detecting zero-day Android malware using the Drebin dataset, which contains over 123,000 benign apps and 5,500 malicious ones. Instead of relying on traditional binary analysis, the approach treated the problem like a natural language task. Features such as opcodes, permissions, and APIs were processed using Google’s BERT transformer, and then classified with a combination of Convolutional Neural Networks (CNNs) and Multi-Layer Perceptrons (MLPs). APKTool was used to extract these features, though no code was publicly released. The results were striking. In zero-day settings, the multi-view deep learning model outperformed individual models with an F1 score of 96\%. With a strong F1 score of 93\%, Random Forest turned out to be the most dependable of the conventional methods. This is consistent with results from the ZDI analysis, which showed that Random Forest was a high performer with about 95\% accuracy. Furthermore, the success of a 1D-CNN in analyzing vulnerability descriptions suggests that treating sequential security data as text is an effective strategy across multiple domains. 
\item From zero-shot machine learning to zero-day attack detection: This study looked at detecting zero-day attacks using the massive UNSW-NB15 and NF-UNSW-NB15-v2 network intrusion datasets, which together contain over 2.5 million samples. The researchers built a Zero-Shot Learning (ZSL) framework in Python with scikit-learn, using Random Forest and Multi-Layer Perceptron (MLP) models. The goal was to connect network features to semantic attributes so the models could recognize attack types they hadn’t seen before. The results were impressive but also highlighted some challenges. The MLP model did particularly well, achieving an average zero-day detection rate of 92.45\% on NetFlow data, outperforming Random Forest. However, performance wasn’t consistent across all attack types for instance, detection for Fuzzers sometimes fell below 20\%. By contrast, supervised learning on ZDI data was much more stable, with models consistently hitting 93–98\% accuracy. This shows that predicting known severity scores from text is inherently more reliable than trying to catch completely new network attacks, even when using similar MLP and Random Forest models. 
\item Use of Data Visualisation for Zero-Day Malware Detection: This paper proposed a hybrid visualization approach to malware detection, working with a massive dataset of 75,000 samples from the VX Heavens database. The researchers converted code behavior into similarity matrices using measures like Euclidean distance and then fed these into Support Vector Machines (SVM) in Python. The Sequential Minimal Optimization (SMO) algorithm produced the greatest results, with a true positive rate of 98.6\% and false positives of less than 2\%, a remarkable standard. Interestingly, this is consistent with the good performance shown in ZDI models. While the malware study used SVMs, ZDI obtained comparable accuracy (95-96\%) with Logistic Regression and ensemble approaches. There is also an interesting parallel in approach: both studies used dimensionality reduction to make complex data easier to comprehend. In the malware work, this involved generating images for classification, which is conceptually similar to how PCA and SVD were applied in ZDI to visualize text features and see how well classes could be separated. 
\item The Performance of Machine and Deep Learning Classifiers in Detecting Zero-Day Vulnerabilities: The study conducted a thorough comparison of 34 classifiers using the Kaggle Microsoft Malware Classification and Meraz'18 datasets to establish the best model for zero-day vulnerability detection. Utilizing Anaconda with Python 3.7 and TensorFlow 1.14, a wide range of algorithms were implemented from Gaussian Naive Bayes to Deep Learning (MLP), ultimately concluding that Random Forest was the superior classifier with 99.51\% accuracy, followed closely by Decision Trees at 99.24\%. Interestingly, they argued that Deep Learning models were computationally expensive for competitive but not superior results (MLP reached 99.33\%), a finding that partially contrasts with the ZDI analysis. The ZDI work showed that Deep Learning architectures like CNNs and LSTMs could match these high-performance levels (~94-95\% accuracy) without requiring excessive tuning, even though Decision Trees were also found to be highly effective (reaching ~98\% accuracy). This is probably because natural language features contain more information than their tabular data. 
\item A framework for detecting zero-day exploits in network flows: This study looked at detecting network flow anomalies using real industrial data from IBM QRadar, along with the public NSL-KDD dataset, and proposed a hybrid framework. Their approach combined a supervised 1D-CNN for feature extraction with unsupervised K-Means clustering and boosting algorithms like Decision Trees and Random Forest, all implemented in Python and TensorFlow. The framework performed impressively, achieving an overall online learning accuracy of 98.4\% on industrial data and 96.6\% on NSL-KDD. Even the CNN component on its own delivered an outstanding F1 score of 0.99. These results reinforce the architectural choices made in the ZDI analysis, where a 1D-CNN implemented in TensorFlow/Keras also achieved very high metrics. Furthermore, there is a clear synergy in design philosophy between the use of stacking classifiers in ZDI to lower false positives in risk scoring and the use of boosting techniques to improve predictions. 
\item Detecting Zero-Day Intrusion Attacks Using Semi\-Supervised Machine Learning Approaches: This research explored several datasets including UNSW-NB15, CICDDoS2019, IoTID20, and CIRA-CIC-DoHBrw-2020 with a focus on picking out the most meaningful features using Benford’s Law. The team built a model that combined this clever feature selection with semi-supervised methods like One-Class SVM (OCSVM) and Gaussian Mixture Models (GMM). Out of all the approaches they tried, OCSVM came out on top, achieving an F1 score of 85\% and a Matthews Correlation Coefficient of 74\%. What’s especially interesting is that the features chosen through Benford’s Law consistently outperformed the ones originally provided with the datasets. It is evident that semi-supervised anomaly detection is a more difficult problem when compared to the supervised models employed in the ZDI analysis, which achieved a Macro-F1 score of almost 92\%. Nevertheless, both studies emphasize the same crucial idea: selecting the appropriate traits is crucial. Just as Benford’s Law helped improve results here, methods like Chi-Square and Mutual Information were essential in the ZDI models for picking out the most important keywords for classification. 
\item Deep Learning for Zero-day Malware Detection and Classification: A Survey: In order to assess deep learning-based zero-day detection, this survey examines a number of datasets from Windows, Android, and hybrid malware contexts. CNNs, RNNs, autoencoders, and adversarial-resistant hybrid architectures are among the models that are covered. The findings highlight patterns demonstrating that deep neural models perform better than traditional machine learning in identifying malware variants that weren't detected before. The ZDI dataset, on the other hand, reflects a parallel trend toward data-driven automation in cybersecurity by using machine learning to the discovery of unknown vulnerabilities rather than malware classification, thereby conceptually aligning. 
\item An Improved Vulnerability Exploitation Prediction Model with Novel Cost Function and Custom Trained Word Vector Embedding: The dataset from National Vulnerability Database (NVD) consists of Text descriptions, exploit status, and severity rankings. To lessen class imbalance, the model incorporates a locally trained word-vector embedding and a proprietary neural network that is improved via a novel cost function. With an accuracy of 0.92 and an F1-score of 0.94, the results reduced overfitting and beat existing methods. By combining TF-IDF with PCA for interpretability, the ZDI dataset, on the other hand, employs a similar methodology but focuses on vulnerabilities from 2024 and expands the predicting power to zero-day vulnerabilities. 
\item Patching zero-day vulnerabilities: an empirical analysis: The dataset includes 4,897 zero-day vulnerabilities collected from 2010–2020 using NVD and the ZDI repository. Based on characteristics such as attack vector, privileges, and confidentiality impact, the model estimates patch release timelines using survival analysis and Cox regression. According to the findings, vulnerabilities that compromise confidentiality take longer to patch than those with a wider scope and vendor impact. Comparatively, the ZDI dataset extends this empirical approach to present risk patterns by supporting comparable survival modeling with contemporary disclosures. 
\item Early and Realistic Exploitability Prediction of Just-Disclosed Software Vulnerabilities: How Reliable Can It Be?: In order to replicate early vulnerability information, the dataset was constructed using NVD and Exploit DB and supplemented with forum conversations from Security Focus and BugTraq. The model evaluated 72 machine learning pipelines, such as pre-trained BERT-based transformers, Random Forest, and Logistic Regression. The findings show that early textual context greatly enhanced exploitability prediction, whereas traditional models such as logistic regression had the highest reliability. This early-warning viewpoint is also included in the ZDI dataset, which, in contrast, focuses on categorizing high-severity vulnerabilities using textual and structured data rather than context acquired from forums. 
\item A Review of Machine Learning-Based Zero-Day Attack Detection: Challenges and Future Directions: This work reviews methods across several datasets, such as CIC-IDS2017 and CSE-CIC-IDS2018 and provides a comprehensive overview rather than a single experimental investigation. It looks at a variety of architectures, from unsupervised autoencoders and GANs to supervised and transfer learning methods, highlighting important issues including the absence of labeled training data for zero-day assaults. The analysis comes to the conclusion that although machine learning has potential, many current techniques have uneven recall and accuracy across various attack types. This high-level perspective contrasts with the ZDI project. While Guo, \cite{guo2023review} emphasizes the industry-wide difficulty of detecting zero-day attacks due to limited labeled data, the use of a rich, retrospective dataset of disclosed vulnerabilities enables the application of high-precision supervised algorithms, such as Random Forest and Decision Trees, to assess severity, achieving consistently high performance (~95–98\%)—a level of stability that is often difficult to attain in real-time detection scenarios highlighted by Guo, \cite{guo2023review}. 
\item Utilising Deep Learning Techniques for Effective Zero-Day Attack Detection: This study explored how to detect zero-day attacks using the CICIDS2017 and NSL-KDD datasets. The researchers built an Autoencoder, a type of deep learning model, and compared it with a One-Class Support Vector Machine (OCSVM). The Autoencoder stood out, achieving detection rates of 89–99\% on NSL-KDD and 75–98\% on CICIDS2017, clearly outperforming the OCSVM and showing its strength in spotting complex, previously unseen attacks. The approach differs from ZDI’s supervised methods, which rely on labeled vulnerability descriptions, whereas the Autoencoder was trained only on “normal” traffic to identify anomalies. Yet, both demonstrate the flexibility of deep learning: just as the Autoencoder captured subtle patterns in network behavior, CNN and LSTM models in ZDI were able to uncover intricate patterns in textual vulnerability data—both achieving impressive accuracy in their respective domains.
\end{itemize}
\section{FEATURE ENGINEERING}

\begin{itemize}
    \item Imported all the necessary libraries for feature engineering
    \item Assigned the dataset columns to shorter variable names for easier reference in later analysis.
    \item Defined two functions to generate a boolean feature indicating the presence of a CVE ID, and to create binary features for detecting specific vulnerability related keywords within the description text.
    \item Performed feature engineering by creating a new binary feature to indicate the presence of a CVE ID, generated keyword indicator features from vulnerability descriptions, added them to the main dataset, and visualized the top 10 most frequent vulnerability-related keywords through a horizontal bar chart.
    \item Converted the CVSS column to numeric values while handling errors, removed entries with missing CVSS scores, created a binary target variable indicating high-severity vulnerabilities (CVSS $\ge 7.0$), defined the feature matrix with descriptive, categorical, and keyword-based features, and displayed the total sample count along with the distribution of high and low CVSS scores.
\end{itemize}
\textbf{TF-IDF + TRUNCATED SVD:}
\begin{itemize}
    \item Built two pipelines: a baseline TF-IDF + Logistic Regression model and a TF-IDF + SVD (100 components) + Logistic Regression variant. Both used a ColumnTransformer combining vendor one-hot encoding, CVE or keyword indicators, and text features. Models were trained on X\_train and y\_train and evaluated on X\_test using classification reports.
    \item Both models performed almost the same, with a total accuracy of about 95.18\% and significant precision and recall for vulnerabilities of high severity (class 1). While dimensionality was lowered, the SVD-reduced model kept the same accuracy, demonstrating that dimensionality reduction enhanced efficiency and interpretability without compromising model performance.
    \item The baseline model's coefficients were analyzed to identify the most influential features. Feature names from the ColumnTransformer were first cleaned and the top 10 positive and negative coefficients were selected. Positive coefficients indicate features associated with a higher likelihood of High CVSS ($\ge 7.0$), while negative coefficients indicate association with lower severity, and magnitudes reflect importance in log-odds terms.
    \item Keywords like “privilege escalation” and “RCE” indicate high severity, while “info disclosure” and “XSS” relate to lower severity, showing clear risk patterns.
    \item Extracted the fitted TF-IDF vectorizer and TruncatedSVD objects from the trained pipeline, retrieved the vocabulary terms, and printed the top 10 most influential words for each of the first 10 SVD components to interpret how textual patterns contribute to the reduced feature space.
    \item Calculated the cumulative explained variance ratio from the fitted TruncatedSVD model and visualized it using a line plot, showing how much total variance in the TF-IDF features is retained as the number of SVD components increases, helping assess the effectiveness of dimensionality reduction.
\begin{figure}[!ht]
    \centering
    \includegraphics[width=0.5\linewidth]{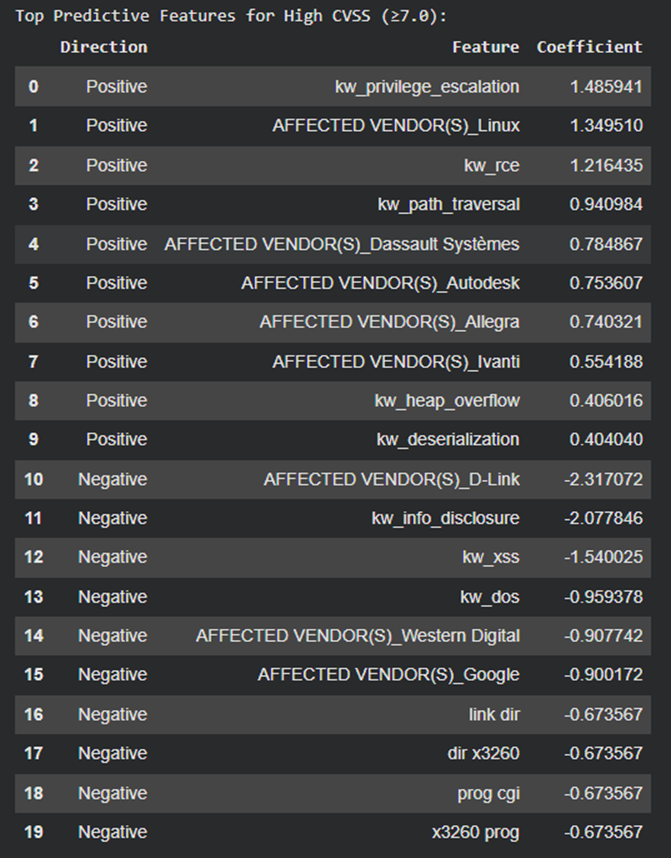}

\end{figure}
\end{itemize}
\textbf{TF-IDF + Chi-Square:}
\begin{itemize}
    \item Developed the TF-IDF + Chi-Square model to identify terms most associated with vulnerability severity. The Chi-Square test measures the strength of correlation between each word feature and the target class, helping to remove less relevant terms. By combining vendor, CVE, and keyword attributes, the model achieved strong precision and recall, showing that Chi-Square effectively selects features that influence classification outcomes.
    \item To understand which text features were most strongly related to high CVSS scores, the Chi-square test was applied to rank n-grams by their importance. The following code extracts and visualizes the top 20 terms with the highest Chi-square values.
\begin{figure}[!ht]
    \centering
    \includegraphics[width=0.75\linewidth]{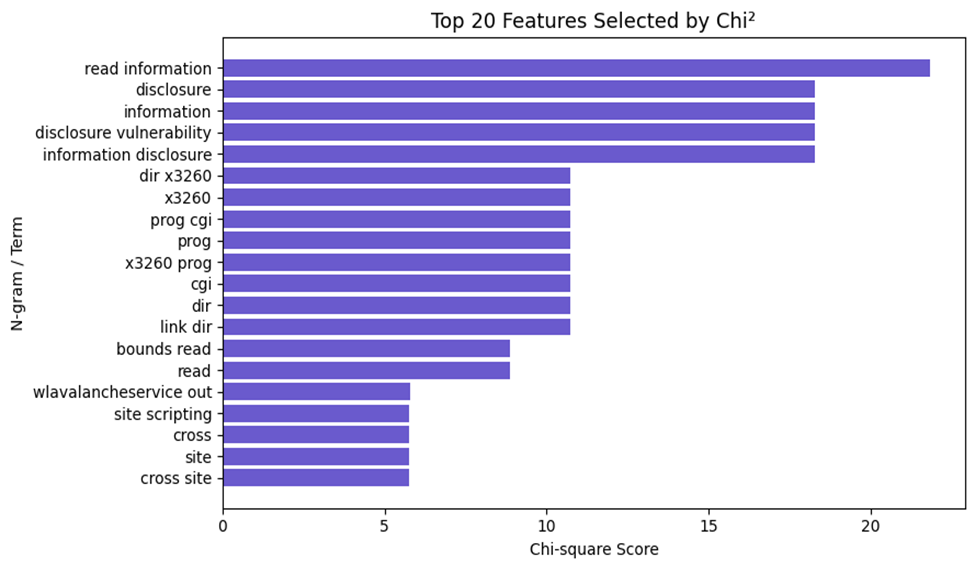}
\end{figure}
\end{itemize}
\textbf{TF-IDF + Mutual Information:}
\begin{itemize}
    \item Built the TF-IDF + Mutual Information model to find which words and features provide the most insight into predicting vulnerability severity. Mutual Information measures how much a feature reveals about the target, enabling the model to capture deeper relationships between text patterns and severity. Using the top 300 features, it performed well in identifying high-severity vulnerabilities.
    \item The model achieved around 95\% accuracy, indicating strong predictive performance. Mutual Information helped highlight features that share meaningful relationships with vulnerability severity.
    \item The fitted TF-IDF vectorizer and Mutual Information selector were extracted from the trained model to identify the most influential text features. The top 20 terms with the highest Mutual Information scores were visualized to illustrate their contribution toward predicting high-severity vulnerabilities.
\begin{figure}[!ht]
    \centering
    \includegraphics[width=0.75\linewidth]{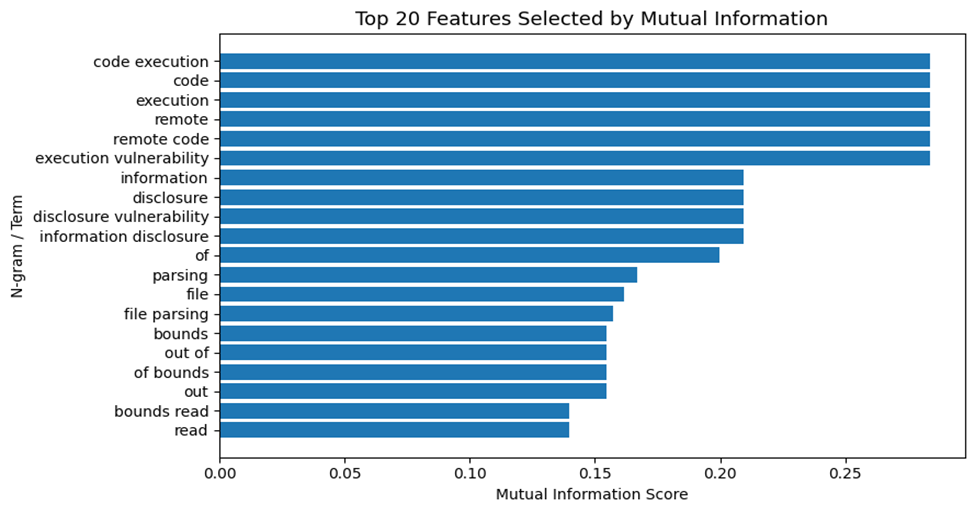}
\end{figure}
    \item The top Mutual Information features, such as “code execution,” “remote code,” and “information disclosure,” highlight patterns strongly linked to high-severity vulnerabilities. These terms reflect common exploit types, suggesting that vulnerabilities involving remote code execution or data disclosure are key indicators of severe security risks.
\end{itemize}
\textbf{PCA \& LDA Feature Engineering}
\begin{itemize}
    \item To cut through the noise and reveal the real story inside the data, we built PCA and LDA pipelines. Think of PCA as providing the ‘big picture’, it maps out the major trends by focusing on where the data varies the most. LDA, on the other hand, acts like a wedge that forces the different groups apart, giving the model a clear, distinct line to separate low-risk issues from high-severity vulnerabilities.
    \item The pipeline first cleans and prepares the data categorical features are encoded, numerical ones are kept as-is, and text is transformed using TF-IDF and Truncated SVD. A small custom step ensures the sparse outputs become dense so the models can handle them easily.
    \item After shrinking the feature space with PCA or LDA, Logistic Regression classifies the data, with classification reports showing its performance. For PCA, a simple 2D scatter plot gives a clear, visual sense of how well the classes are separated, making it easy to see the structure of the data.
\end{itemize}
\begin{figure}[!ht]
    \centering
    \includegraphics[width=0.75\linewidth]{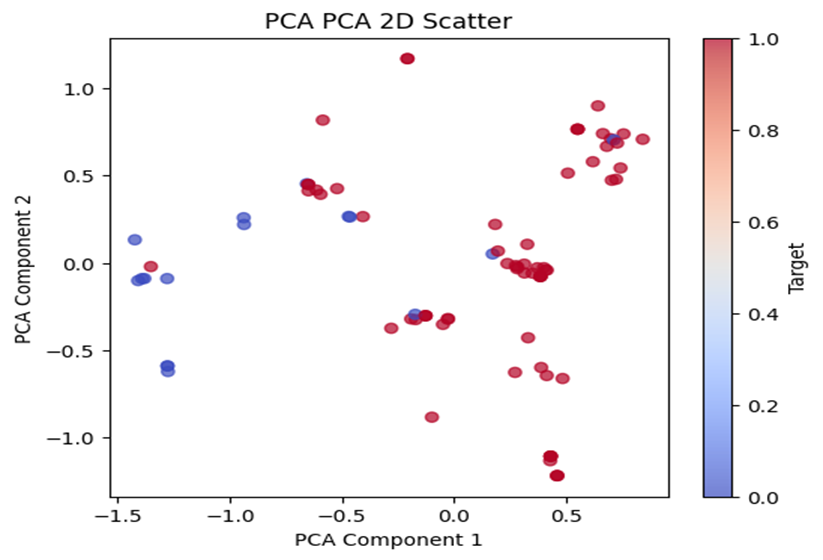}
\end{figure}

\begin{table*}[t]
\centering
\caption{Performance Comparison of Feature Selection and Dimensionality Reduction Techniques}
\label{tab:feature_selection}
\resizebox{\textwidth}{!}{%
\begin{tabular}{l p{3.5cm} c p{4.5cm} c c c}
\toprule
\textbf{Model (Technique)} & \textbf{Feature Type} & \textbf{Classifier} & \textbf{Key Strength} & \textbf{Accuracy} & \textbf{Macro-F1} & \textbf{ROC-AUC} \\
\midrule
TF-IDF Baseline & Text, vendor, CVE, and keyword features & LR & Gave steady and reliable results without any feature reduction & 0.9518 & 0.9186 & \textbf{0.9944} \\
TF-IDF + Truncated SVD & Condensed text features capturing main patterns & LR & Kept accuracy stable while reducing feature count & 0.9518 & 0.9186 & 0.9935 \\
TF-IDF + Chi-Square & Selected text features based on importance to output & LR & Focused on the most relevant words affecting prediction & 0.9518 & 0.9141 & \textbf{0.9944} \\
TF-IDF + Mutual Info. & Text features strongly related to severity & LR & Balanced accuracy with simpler, meaningful features & 0.9518 & 0.9186 & 0.9925 \\
PCA & Principal components maximizing variance & LR & Achieved highest accuracy by capturing global structure & \textbf{0.9639} & \textbf{0.9373} & 0.9916 \\
LDA & Linear discriminants maximizing separability & LR & Focused on axes that best separate severity classes & 0.9518 & 0.9226 & 0.9216 \\
\bottomrule
\end{tabular}%
}
\end{table*}

\section{MACHINE LEARNING MODELS}

\textbf{1. }\textbf{Classical models}

Four models were selected:

\begin{itemize}
    \item \textbf{Logistic Regression} – A simple linear model that works well as a baseline, especially for binary classification tasks.
\end{itemize}

\begin{itemize}
    \item \textbf{Random Forest} – An ensemble of decision trees that helps capture non-linear patterns and reduces overfitting through bagging.
\end{itemize}

\begin{itemize}
    \item \textbf{Decision Tree} – A single decision tree model that is easy to interpret and useful for understanding decision boundaries.
\end{itemize}

\begin{itemize}
    \item \textbf{K-Nearest Neighbors} – A distance-based classifier that predicts labels by looking at nearby samples in the feature space.
\end{itemize}

Before the model was trained, each classifier was put inside a Scikit-learn pipeline, where all feature transformations were handled by the specified preprocessing block (pre). By using the same preprocessing procedures for both training and testing, this method guaranteed consistency throughout all experiments and avoided data leaking.

Training was carried out on the training split of the dataset, which included TF-IDF vectorized descriptions along with structured metadata. This allowed the classifiers to learn from both textual information and contextual fields present in the dataset.

After training, each pipeline generated predictions on the unseen test set to assess generalization performance. This ensured that all evaluation results reflected model behavior on data not used during training.

For every model, a classification report was produced, summarizing accuracy, precision, recall, and F1-score for each class. These metrics provided a detailed comparison of how effectively each classifier distinguished between the two severity categories.

\begin{figure}[H]
    \centering
    \begin{minipage}{0.25\linewidth}
        \includegraphics[width=\linewidth]{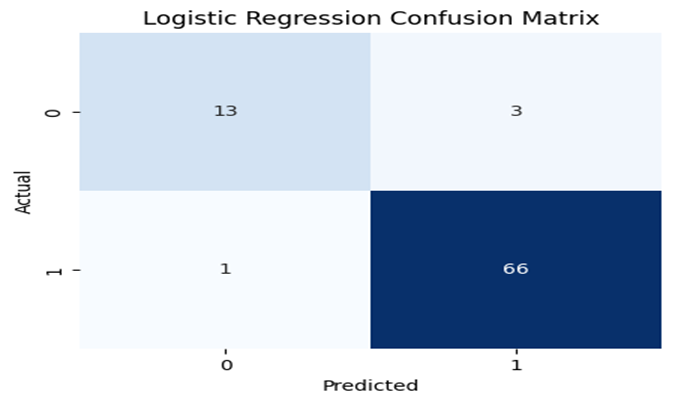}
    \end{minipage}%
    \begin{minipage}{0.25\linewidth}
        \includegraphics[width=\linewidth]{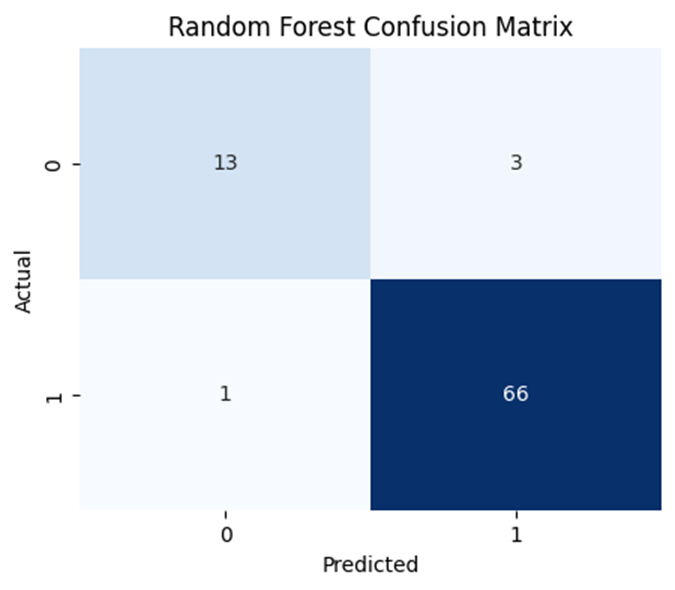}
    \end{minipage}%
    \begin{minipage}{0.25\linewidth}
        \includegraphics[width=\linewidth]{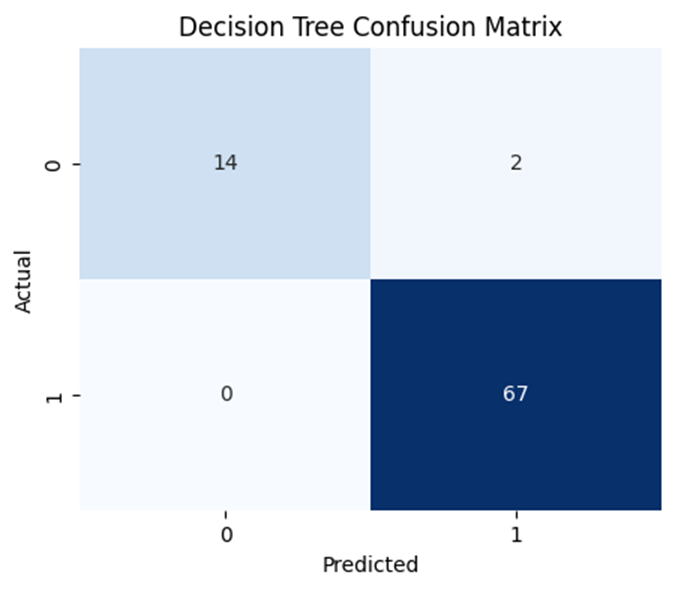}
    \end{minipage}%
    \begin{minipage}{0.25\linewidth}
        \includegraphics[width=\linewidth]{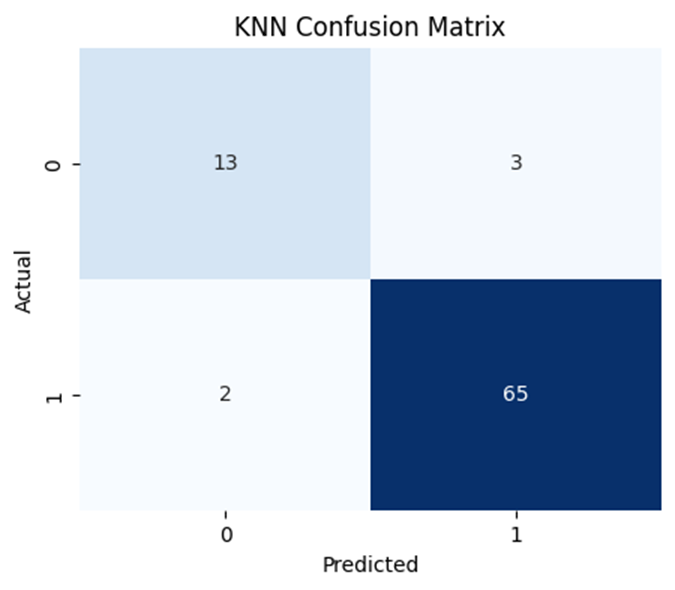}
    \end{minipage}
\end{figure}

\textbf{2. }\textbf{Deep learning models}

Three models were selected namely Feed Forward Neural Network(FFNN), Long Short-Term Memory(LSTM) and Convolutional Neural Network(CNN).

Then the code was developed that pre-processes the dataset by converting text descriptions into padded integer sequences for the LSTM branch of the model. At the same time, categorical and numerical metadata are transformed using a ColumnTransformer to create the tabular input used by the FFNN branch. The text and tabular features are kept separate, and their input dimensions are calculated for building the combined multi-input deep learning architecture.

\textbf{a) }\textbf{Feed Forward Neural Network(FFNN)}

The model is designed with two complementary branches, so it can learn from both text and structured data at the same time. The text is first turned into meaningful numerical representations using an Embedding layer, and then a global average pooling step squeezes this information into a simple, easy-to-work-with vector. In parallel, the numerical and categorical features go through their own small dense layer, which helps the model pick up useful patterns in the tabular data. Once both branches have done their part, their outputs are combined and passed through a few fully connected layers with dropout. This helps the model to generalize better before it finally produces a single sigmoid output for the binary prediction.

After training, the focus shifts to understanding what the model has actually learned. To make this clearer, an intermediate version of the model is created to extract the activations from the 64-neuron dense layer. By visualizing these activations as a heatmap for a sample test input, it becomes much easier to see which neurons are firing strongly. This offers an intuitive glimpse into how the model mixes information from the text and the tabular features to arrive at its final decision.
\begin{figure}[!ht]
    \centering
    \includegraphics[width=0.5\linewidth]{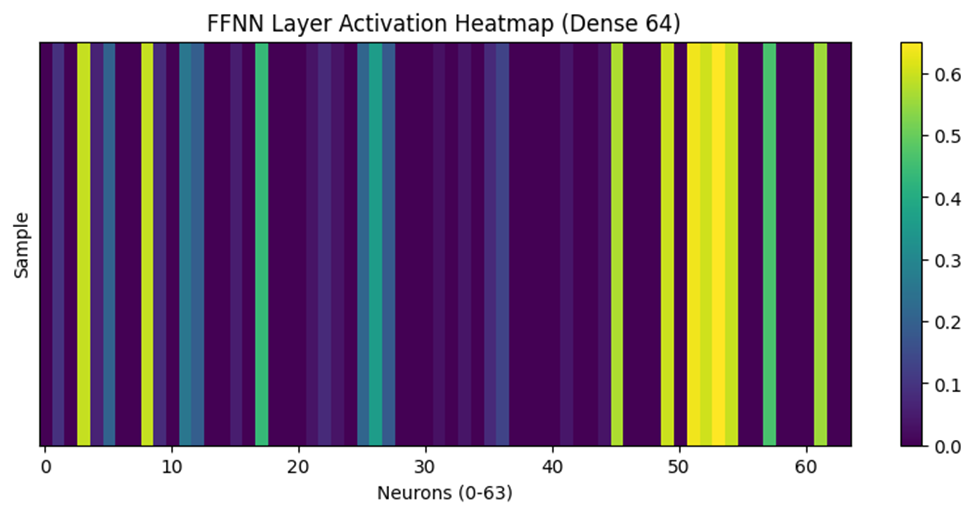}

\end{figure}

\textbf{b) }\textbf{Long Short-Term Memory(LSTM)}

A two-headed LSTM model is defined with separate inputs for text and tabular data. The text input is passed through an Embedding layer and then through an LSTM layer, which summarizes the sequence into a fixed-length vector. The tabular input, which includes numerical and categorical information, is processed via a tiny dense layer, and the outputs of both heads are concatenated and passed through additional dense layers with dropout, resulting in a single sigmoid output for binary classification.

The model is trained using both text and tabular inputs simultaneously on the training dataset, with validation on a held-out split. Following training, an intermediate model is developed to extract the activations of the LSTM layer for a test sample. These activations are then displayed as a heatmap to show which hidden units are most active, giving information about how the LSTM encodes text before merging it with metadata for classification.

\begin{figure}[!ht]
    \centering
    \includegraphics[width=0.5\linewidth]{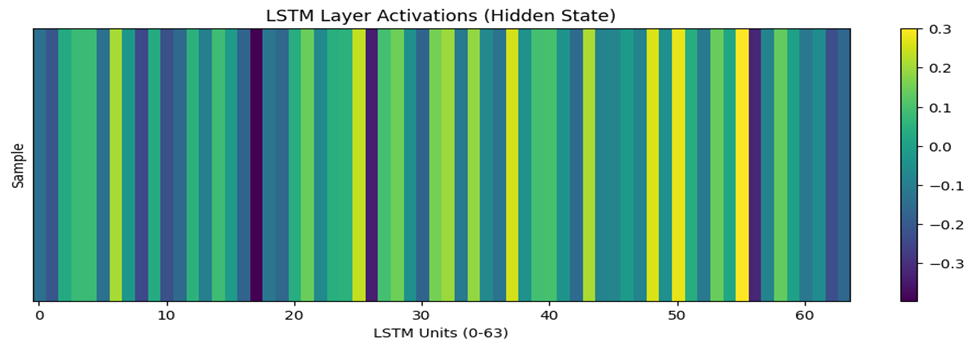}

\end{figure}
\textbf{c) }\textbf{Convolutional Neural Network(CNN)}

Text and tabular data are defined as distinct inputs for a two-headed CNN model. After passing the text input through an Embedding layer and a Conv1D layer, the sequence is scanned for local patterns, such as n-grams. After that, each filter's strongest feature is extracted using global max pooling, and the tabular input is processed through a slightly dense layer. The outputs of both heads are concatenated and further passed through dense layers with dropout to produce a single sigmoid output for binary classification.

The model is trained using both text and tabular inputs simultaneously, with validation on a held-out split. An intermediate model is created to extract the feature maps of the Conv1D layer for a test sample, and the first eight filters are visualized, showing which parts of the text each filter focuses on. This helps interpret how the CNN captures local patterns in the text before combining them with metadata for prediction.

\begin{figure}[!ht]
    \centering
    \includegraphics[width=0.5\linewidth]{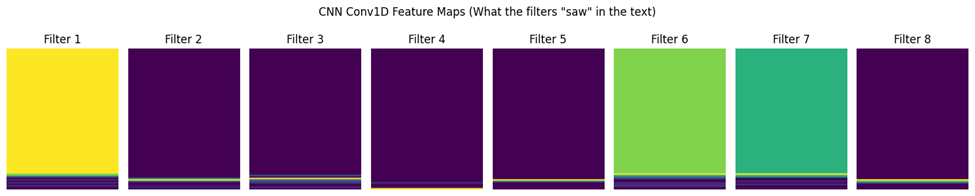}

\end{figure}

\textbf{Deep learning models comparison}

A unified evaluation function is used for all deep learning models, taking the model, test inputs, and true labels as inputs. The function first predicts probabilities for the test samples and converts them into binary labels using a threshold of 0.5. It then generates a detailed classification report showing accuracy, precision, recall, and F1 score, providing a clear view of each model’s performance on the test set.

In addition, the function visualizes the confusion matrix to highlight correct and incorrect predictions and plots the ROC curve using the predicted probabilities to evaluate how well the models can distinguish between the two classes. This standardized evaluation approach is applied to the Feed-Forward Neural Network, LSTM, and CNN models, ensuring a fair and consistent comparison across all architectures.

\begin{figure}[H]
    \centering
    \begin{minipage}{0.25\linewidth}
        \includegraphics[width=\linewidth]{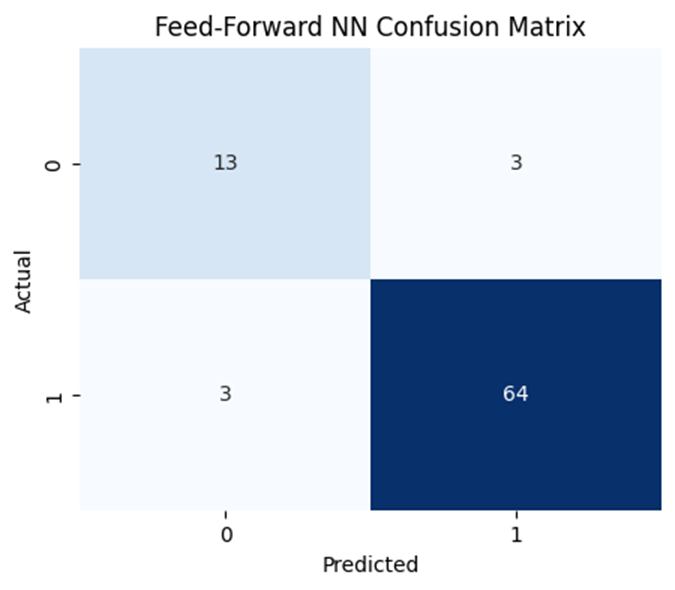}
    \end{minipage}%
    \begin{minipage}{0.25\linewidth}
        \includegraphics[width=\linewidth]{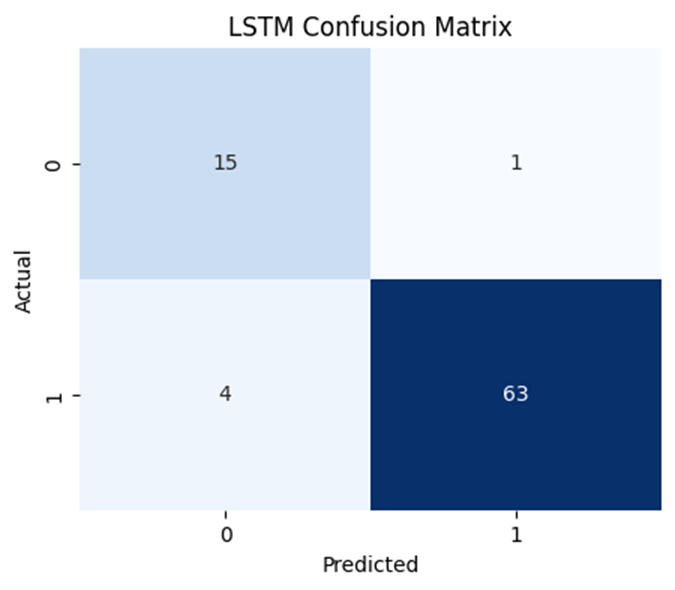}
    \end{minipage}%
    \begin{minipage}{0.25\linewidth}
        \includegraphics[width=\linewidth]{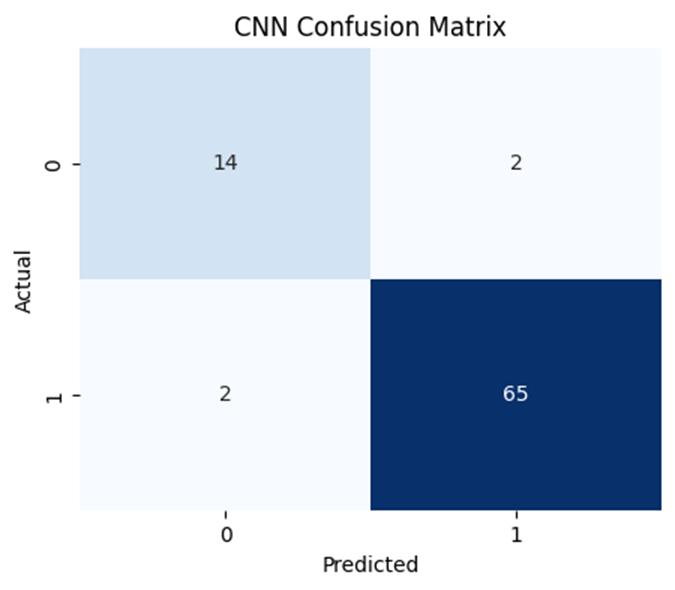}
    \end{minipage}
\end{figure}

\textbf{Overall Machine Learning Models Comparison}

Standard evaluation measures, such as accuracy, macro precision, macro recall, macro F1-score, and ROC-AUC where probabilities are known, are computed using a helper function. In order to fit each pipeline on the training data, produce predictions, publish the classification report, and save the metrics in a list for further comparison, the algorithm first loops over all of the classical models.

The deep learning models Feed-Forward NN, LSTM, and CNN are evaluated similarly using the same metrics. Predictions and probabilities are collected, reports are printed, and all results from classical and deep learning models are combined into a single DataFrame, providing a unified benchmark table for easy comparison of model performance across architectures.

\begin{table*}[t]
\centering
\caption{Final Benchmark of Classical vs. Deep Learning Models (Support: 83 samples). Best scores are bolded.}
\label{tab:final_benchmark}
\resizebox{\textwidth}{!}{%
\begin{tabular}{l c c c c c}
\toprule
\textbf{Model} & \textbf{Accuracy} & \textbf{Precision (Macro)} & \textbf{Recall (Macro)} & \textbf{Macro-F1} & \textbf{ROC-AUC} \\
\midrule
Classical: Logistic Regression & 0.9518 & 0.9425 & 0.8988 & 0.9186 & \textbf{0.9944} \\
Classical: Random Forest & 0.9518 & 0.9425 & 0.8988 & 0.9186 & 0.9921 \\
Classical: Decision Tree & \textbf{0.9759} & \textbf{0.9855} & 0.9375 & \textbf{0.9593} & 0.9375 \\
Classical: KNN & 0.9398 & 0.9113 & 0.8913 & 0.9008 & 0.9548 \\
\midrule
Feed-Forward NN & 0.9277 & 0.8839 & 0.8839 & 0.8839 & 0.9757 \\
LSTM & 0.9398 & 0.8869 & \textbf{0.9389} & 0.9095 & 0.9925 \\
CNN & 0.9518 & 0.9226 & 0.9226 & 0.9226 & 0.9935 \\
\bottomrule
\end{tabular}%
}
\end{table*}

\begin{figure}[!ht]
    \centering
    \includegraphics[width=0.5\linewidth]{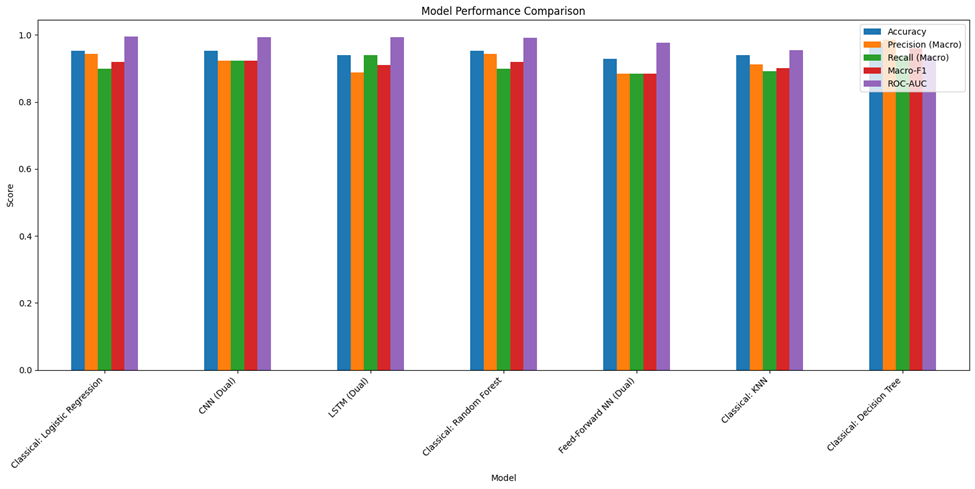}
\end{figure}
\textbf{3. }\textbf{ENSEMBLE MODELS}

 \textbf{a) }\textbf{Feature based ensemble}

Two different logistic regression models are trained: While Model B just uses numerical and categorical features, Model A combines text features (TF-IDF of descriptions) with keywords and structured metadata. A pipeline that applies the required preprocessing is used to fit each model, guaranteeing uniform treatment of textual, numerical, and categorical inputs.

An ensemble prediction is developed by averaging the probability outputs of both models' predictions. Accuracy and a classification report are used to evaluate the overall efficacy of an ensemble technique, which makes use of complimentary information from text and structured features.

\begin{figure}[!ht]
    \centering
    \includegraphics[width=0.5\linewidth]{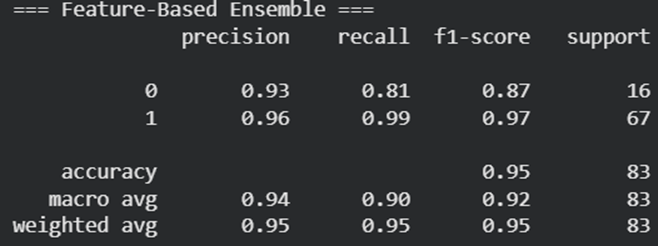}
\end{figure}
\textbf{b) }\textbf{Data based ensemble}

Different bootstrapped subsets of the training data, each comprising 80\% of the original samples, are used to train five different logistic regression models. By introducing variation to the training sets, resampling lowers the chance of overfitting and enables each model to detect slightly different patterns.

The final ensemble probability for each test sample is calculated by taking the average of the five models' anticipated results. A 0.5 threshold is then used to transform the final ensemble probability into binary labels. By combining several models trained on different subsets of data, this data-based ensemble is assessed using accuracy and a classification report, offering a reliable evaluation of performance.

\begin{figure}[!ht]
    \centering
    \includegraphics[width=0.5\linewidth]{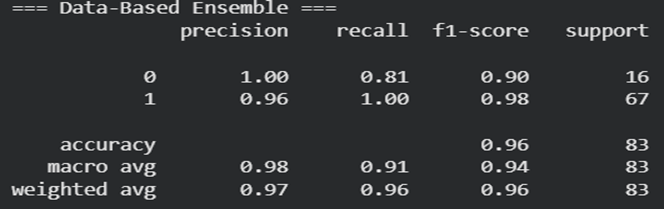}
\end{figure}
\textbf{c) }\textbf{Model based ensemble}

Using a uniform pre-processing pipeline, three distinct classical classifiers Logistic Regression, Random Forest, and K-Nearest Neighbors are each trained on the same training set. For the test set, each model produces probability predictions that capture various learning biases and decision-making techniques.

All three models first make their own probability guesses, then these are simply averaged, and anything above 0.5 is treated as class 1 while anything below is class 0. The ensemble’s quality is then checked using accuracy and a classification report, which together show how well combining the models smooths out their individual weaknesses and leads to more dependable predictions.

\begin{figure}[!ht]
    \centering
    \includegraphics[width=0.5\linewidth]{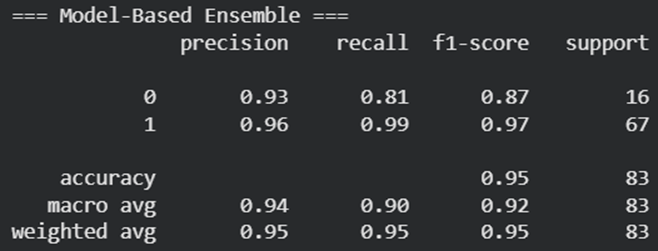}
\end{figure}
\textbf{d) }\textbf{Model Instance based ensemble}

Three separate Logistic Regression models are trained on the same data, each with a different regularization strength (C = 0.5, 1.0, 2.0), so that they learn slightly different decision boundaries while using the same features and pre-processing. The probability outputs of these three models are then averaged, and anything above 0.5 is treated as class 1 (otherwise class 0); this model-instance ensemble, evaluated using accuracy and a classification report, aims to smooth out individual model quirks and produce more stable, reliable predictions.

\begin{figure}[!ht]
    \centering
    \includegraphics[width=0.5\linewidth]{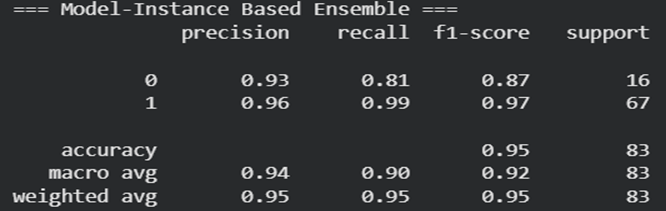}
\end{figure}
\textbf{e) }\textbf{Output based ensemble}

A stacking ensemble is built with three base models Logistic Regression, Random Forest, and K-Nearest Neighbors, whose predictions are fed as inputs into a final Logistic Regression “meta” model that learns how to best blend their outputs. Predictions are made on the test set once this end-to-end pipeline has been trained on the preprocessed training data. Metrics like accuracy and the classification report are used to demonstrate how well the ensemble integrates these various classifiers into a single, trustworthy predictor.

\begin{figure}[!ht]
    \centering
    \includegraphics[width=0.5\linewidth]{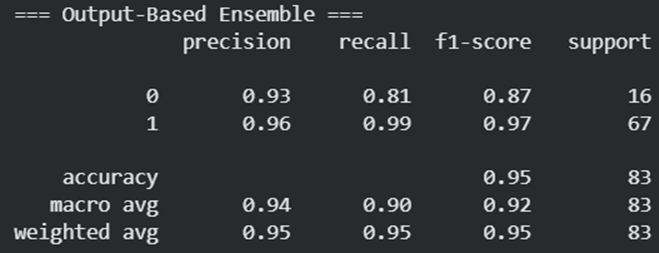}
\end{figure}

\section{COMPARISON WITH KAGGLE CODES}

The table compares the ZDI analysis project with several Kaggle-style vulnerability and cybersecurity studies, highlighting differences in goals, datasets, feature types, and models. The ZDI project uses a dual-input technique (text embeddings and vendor metadata) with hybrid models including CNN, LSTM, Decision Tree, and Logistic Regression, whilst other research uses AutoEncoders, classical regressors, or RNN-based architectures depending on the dataset and task.

The main findings demonstrate that whereas earlier studies using simpler or single-input models obtain reasonable accuracy, the dual-input, hybrid deep learning strategy delivers good stability and predictive performance (ROC-AUC > 0.99). This shows that a performance benefit over solely classical or text-only techniques can be obtained by integrating textual and metadata information.

\begin{table*}[t]
\centering
\caption{Comparison with kaggle codes}
\label{tab:related_projects}
\resizebox{\textwidth}{!}{%
\begin{tabular}{p{3cm} p{3.5cm} p{4cm} p{3cm} p{4cm}}
\toprule
\textbf{Project Name} & \textbf{Objective} & \textbf{Data \& Approach} & \textbf{Models Used} & \textbf{Key Outcomes} \\
\midrule
\textbf{ZDI Analysis} & Analyze trends in zero-day vulnerabilities and predict high-severity cases (CVSS $> 7.0$). & \textbf{Dataset:} ZDI 2024 \newline \textbf{Features:} Dual-input approach combining sequence text embeddings with one-hot/numeric vendor metadata. & Hybrid CNN, Hybrid LSTM, Decision Tree, Logistic Regression, Random Forest & The Decision Tree hit the highest accuracy ($\sim$97.6\%), though Hybrid CNN \& LR offered better stability (ROC-AUC $> 0.99$). \\
\midrule
\textbf{Celosia Zero-Day} & Detect network anomalies and potential intrusions. & \textbf{Dataset:} N-BaIoT \newline \textbf{Features:} Scaled numerical features derived from IoT network traffic patterns. & AutoEncoder, Logistic Regression & The AutoEncoder successfully isolated attacks from normal traffic, achieving 100\% accuracy on synthetic test data. \\
\midrule
\textbf{CVE Severity Classification (Comparative)} & Classify the severity levels of vulnerabilities. & \textbf{Dataset:} NIST CVE \newline \textbf{Features:} Combined text embeddings with one-hot encoded vendor metadata. & MLP-LSTM, RoBERTa, XGBoost & Hybrid Deep Learning models proved superior to classical algorithms, achieving the best result with $\sim$72\% accuracy. \\
\midrule
\textbf{CVE Severity Classification (DL)} & Use Deep Learning to classify vulnerability severity. & \textbf{Dataset:} Global CVE \& CWE \newline \textbf{Features:} Tokenized text enriched with GloVe word embeddings. & Bi-LSTM, GRU, RNN & The Bi-LSTM model paired with GloVe embeddings came out on top ($\sim$71\% accuracy), effectively capturing sequential context. \\
\midrule
\textbf{Cybersecurity Threat Analysis} & Perform Exploratory Data Analysis (EDA) and map network relationships. & \textbf{Dataset:} CISA 2022 \newline \textbf{Features:} Analysis of vendor-product networks (with imputed missing data). & Bayesian Network, K-Means Clustering, Naive Bayes & Successfully visualized complex vendor links and identified critical ``super-spreader'' vulnerabilities within the ecosystem. \\
\midrule
\textbf{CVSS Prediction} & Predict specific CVSS scores (Regression task). & \textbf{Dataset:} CVE Common Vulnerabilities \newline \textbf{Features:} TF-IDF text analysis combined with one-hot encoded access/impact vectors. & Random Forest Regressor & The model achieved an impressive $R^2$ of 0.99, proving that combining text analysis with vector features is highly effective. \\
\bottomrule
\end{tabular}%
}
\end{table*}

\section{Conclusion}

Zero-day vulnerabilities remain among the most challenging and consequential threats in modern cybersecurity due to the absence of prior knowledge, available patches, and established defensive mechanisms at the time of exploitation. In this paper, we presented a comprehensive empirical analysis of zero-day vulnerabilities disclosed by the Zero Day Initiative (ZDI) between January and April 2024, leveraging a dataset of 415 vulnerability reports that include severity scores, vendor information, and textual descriptions.

Our analysis highlighted notable trends in vulnerability disclosures and severity distributions across vendors, reinforcing the observation that vulnerabilities affecting widely deployed software ecosystems are more likely to exhibit higher CVSS scores and broader potential impact. By incorporating both structured attributes and unstructured textual descriptions, we demonstrated that machine learning models can assist in early-stage severity assessment, even when full contextual information is not yet available.

The comparative evaluation of classical machine learning models and deep learning approaches revealed complementary strengths. Classical models exhibited strong performance when trained on structured features and benefited from dimensionality reduction techniques, while deep learning models, particularly those capable of capturing semantic context in vulnerability descriptions, showed improved recall for high-severity cases. Importantly, our results emphasize that accuracy alone is insufficient for evaluating vulnerability prediction systems. Metrics such as precision, recall, F1-score, and ROC--AUC provided more meaningful insights into model effectiveness under class imbalance, particularly for identifying critical vulnerabilities that warrant immediate attention.

Overall, this study demonstrates the practical value of combining empirical vulnerability analysis with machine learning–based severity prediction to support more informed prioritization and patch management decisions. Future work will extend this analysis to larger temporal windows, incorporate exploit availability and patch latency data, and explore explainable artificial intelligence techniques to improve transparency and trust in automated vulnerability assessment systems.

\bibliographystyle{unsrtnat}
\bibliography{references}

\end{document}